\documentstyle[preprint,aps,12pt]{revtex}
\tightenlines
\begin{document}
\draft
\hfill\vbox{\baselineskip14pt
            \hbox{\bf PCP-65/ISS 2001}
            \hbox{June 2001}}
\baselineskip20pt
\vskip 0.2cm 
\begin{center}
{\Large\bf Quantum Groups, Strings and HTSC materials II}
\end{center} 
\begin{center}
\large Sher~Alam$^{1}$,~M.O.Rahman$^{3}$,~T.~Yanagisawa$^{2}$~and
~H.~Oyanagi$^{1}$
\end{center}
\begin{center}
$^{1}${\it Photonics, Nat.~Inst.~of~AIST, Tsukuba, Ibaraki 305-8568, Japan}\\
$^{2}${\it Nanoelectronics, Nat.~Inst.~of~AIST, Tsukuba, Ibaraki 
305-8568, Japan}\\
$^{3}${\it  GUAS \& Photon Factory, KEK, Tsukuba, Ibaraki 305, Japan}
\end{center}
\begin{center} 
\large Abstract
\end{center}
\begin{center}
\begin{minipage}{16cm}
\baselineskip=18pt
\noindent
	
  Previously we have indicated the relationship between quantum groups 
[Phys. Lett A272, (2000)] and strings via WZWN models. In this note 
we discuss this relationship further and point out its possible 
applications to cuprates and related materials.The connection between 
quantum groups and strings is one way of seeing the validity of our previous
conjecture [i.e. that a theory for cuprates may be constructed on the 
basis of quantum groups]. The cuprates seems to exhibit statistics, 
dimensionality and phase transitions in novel ways. The nature of 
excitations [i.e. quasiparticle or collective] must be understood. 
The Hubbard model captures some of the behaviour of the phase transitions 
in these materials. On the other hand the phases such as stripes in these 
materials bear relationship to quantum group or string-like solutions. 
One thus expects that the relevant solutions of Hubbard model may thus 
be written in terms of stringy solutions. In short this approach may 
lead to the non-perturbative formualtion of Hubbard and other condensed 
matter Hamiltonians. The question arises that how a 1-d based symmetry 
such as quantum groups can be relevant in describing 
a 3-d [spatial dimensions] system such as cuprates. The answer lies in 
the key observation that strings which are 1-d objects can be used to 
describe physics in $d$ dimensions. For example gravity [which is 
a 3-d [spatial] plus time] phenomenon can be understood in terms of 
1-d strings. Thus we expect that 1-d quantum group object induces physics 
in 2-d and 3-d which may be relevant to the cuprates. We present
support for our contention using [numerical] variational Monte-Carlo
[MC] applied to 2d d-p model. We also briefly
discuss others ways to formulate a string picture for cuprates, 
namely by exploiting connection between gauge theories and strings 
and t'Hooft picture of quark confinement.  
\end{minipage}
\end{center}
\vfill
\baselineskip=20pt
\normalsize
\newpage
\setcounter{page}{2}
   One of the central points of our contention is:
\begin{itemize}
\item{}Relevant charge degrees of freedom are 1-d.
\item{}The spin degrees of freedom also exhibit a 1-d behavior. 
\end{itemize} 
Moreover the phonon-electron coupling can be treated in
terms of a 1-d model, such Hubbard Holstein.
The 1-d behavior of magnetic fluctuation was predicted
by our theory \cite{alam99-1} before the experiment!
The fluctuations associated with charge were regarded
as 1-d, whereas magnetic fluctuations were regarded
as 2-d. However it was predicted \cite{alam99-1} and
experimentally shown \cite{moo00} that magnetic
fluctuations are also 1-d. Thus in our scenario
all relevant degrees of freedom are 1-d instead of
quasiparticle in the sense of Landau. Superconductivity
arises as a dressed stripe phase. We think it may be
possible to write down transformations equivalent 
[such as one writes in BCS, i.e. Bogoliubov] which
can map the 1-d states onto the 3-d superconducting
phase. Let us now discuss some of the rationale behind our
thinking.

	In a previous work one of us \cite{alam98} has advanced 
the conjecture that one should attempt to model the phenomena of
antiferromagnetism and superconductivity by using quantum
symmetry group. Following this conjecture to model the phenomenona of
antiferromagnetism and superconductivity by quantum symmetry
groups, three toy models were proposed \cite{alam99-1}, namely,
one based on ${\rm SO_{q}(3)}$ the other two constructed with
the ${\rm SO_{q}(4)}$ and ${\rm SO_{q}(5)}$ quantum groups. 
Possible motivations and rationale for these choices  
were outlined \cite{alam99-1}. In \cite{alam99-2} a model to 
describe quantum liquids in transition from 1d to 2d dimensional 
crossover using quantum groups was outlined. In \cite{alam00-1} 
the classical group ${\rm SO(7)}$ was proposed as a toy model to 
understand the connections between the competing phases and the 
phenomenon of psuedo-gap in High Temperature Superconducting Materials
[HTSC]. Then we proposed in \cite{alam00-2} an idea to
construct a theory based on patching critical points so as to
simulate the behavior of systems such as cuprates.
To illustrate our idea we considered an example 
discussed by Frahm et al., \cite{fra98}. The model
deals with antiferromagnetic spin-1 chain doped with
spin-1/2 carriers. In \cite{alam00-3} the connection between
Quantum Groups and 1-dimensional [1-d] structures such as
stripes was outlined. The main point of \cite{alam00-3}
is to emphasize that {\em 1-d structures play an important
role in determining the physical behaviour [such as the 
phases and types of phases these materials are capable of
exhibiting] of cuprates} and related materials.

	The question arises that how a 1-d based symmetry
such as quantum groups can be relevant in describing a
3-d [spatial dimensions] system such as cuprates. {\em The
answer lies in the key observation that strings which
are 1-d objects can be used to describe physics in
$d$ dimensions}. For example gravity \cite{kak91}
[which is a 3-d [spatial]
plus time] phenomenon can be understood in terms of 1-d
strings. Thus we expect that 1-d quantum group object
induces physics in 2-d and 3-d which may be relevant to
the cuprates. However we like to point out that
the notion of string is more general and does not
have to be restricted to quantum groups. The reason
for our choice are several. For one quantum groups
are intimately connected with WZWN models and
to strings. In turn WZWN based models are relevant
to disordered systems. The exact solution of Hubbard
model in 1-d is governed by quantum group symmetry.
In turn the Hubbard model captures much of the physics
of cuprates. We note that the t-J model and its
various generalizations are also advocated for the study
of cuprates. However the t-J model is but a limit of Hubbard
model [i.e. when $U/t \rightarrow \infty$]. Quantum
critical points may be naturally understood in terms
of quantum groups.

	Let us briefly comment on three different
general aspects of the cuprates and related materials:
\begin{itemize}
\item{}Statistics and fractionalization:-The cuprates seem 
to exhibit exotic statistics and charge fractionalization. 
Indications for electron fractionalization from Angle-Resolved 
Photoemission Spectroscaopy [ARPES] have
been reported. We have suggested to measure fractionalization
by using SET based experiments \cite{alam00-4}
\footnote{After about a week or so later of our submission a paper 
by Sentil and Fisher ``Detecting fractions of electrons in the 
high-$T_c$ cuprates''cond-mat/0011345, appeared without reference to 
our work. Some mathematical details were worked out but it used or 
was in line with basically our idea.}.

We note that the Fermi 
liquid is characterized by sharp fermionic
quasiparticle excitations and has a discontinuity in the
electron momentum distribution function. In contrast
the Luttinger liquid is characterized by charge $e$
spin $0$ bosons and spin $1/2$ charge $0$ and the
fermion is a composite of these [i.e. fractionalization].
It is well-known that transport properties
are defined via correlation functions.
The correlation functions of a Luttinger liquid
have a power law decays with exponents that
depend on the interaction parameters. Consequently
the transport properties of a Luttinger liquid
are very different from that of a Fermi liquid.
Photoemission experiments on Mott insulating oxides
seems to indicate the spinon and holon excitations
of a charge Luttinger liquid. However the experimental
signatures of Luttinger liquid are not totally
convincing. To this end we propose SET based experiments
to determine the Luttinger liquid behaviour of
the cuprates.
\item{}Mixed dimensionality:- The cuprates and related
materials seem to exhibit mixed dimensionality. This
can be seen in many materials. We give an example of
Ca$_{2+x}$ Y$_{2-x}$ Cu$_{5}$ O$_{10}$ \cite{yam99}, a material 
which simple [i.e. simplicity in structural aspect and controlled
hole doping] and exhibits mixed dimensionality and can 
provide insight into the spin and charge dynamics of 
more complicated HTSC material. As a consequence of mixed 
dimensionality this material exhibits ferromagnetism and 
antiferromagnetism, introduction of holes leads to novel 
spin-charge dynamics in this magnetically frustrated system.
\item{}Phases:-A variety of phases have been
reported in cuprates, for example, antiferromagnetic insulator,
superconducting, metallic, strange metal, spin-glass,
and insulating. 
\end{itemize}

	It has been known that the symmetry of Hubbard 
model is $SU(2)\times SU(2)/Z_2 \equiv SO(4)$, this
can be taken as a motivation to consider a gauge
model or NLSM for Hubbard based on $SU(2)$, $SU(2)\times SU(2)$,
$SO(4)$ or even $Z_2$ or many combinations and
generalizations thereof depending on one's point
of view. We take the following point of view.
It is known that quantum groups in the context of
usual HH are tied to 1-d. This is used as an argument
against the application of quantum group to higher 
dimensions. We take the opposite attitude. Using the 
string notion instead of going from 1-d, to 2-d via
the usual route, we assume that fundamental
entities are 1-d objects [subject to quantum
group symmetry] i.e. strings, for example we introduce
the notion of Hubbard string. And attach the known
symmetery of Hubbard Hamiltonian i.e. $SU(2)\times 
SU(2)/Z_2 \equiv SO(4)$ for example as a Chan-Paton 
factor\footnote{Chan-Paton factor:-An open string has 
boundaries, i.e. endpoints, in quantum with distinguished 
endpoints it is natural to associate degrees of freedom
with these in addition to fields propagating in bulk.
Moreover it is natural and necessary to have simply
boundary conditions. For example in string theory
of strong interactions, one introduces SU(3)
flavor quantum numbers, the endpoints have 
quark-antiquark attached connected by `color-electric
flux tube}. We conjecture that the solution of 2-d
and 3-d Hubbard Hamiltonian can be written in
terms of the 1-d Hubbard string solutions.
This conjecture can be tested by looking at
specific examples. We thus have provided
a general framework for strongly correlated
electron systems, such as Quantum Hall systems
and HTSC materials and related materials. The recent 
data supports our earlier conjecture of important 
role of 1-d systems tied to quantum groups. 

P.W.Anderson and others have tried to prove spin charge
separation in 2-d. In our framework this is
not necessary and follows naturally as a 
consequence of the Hubbard string. It is
clear that spin charge separation takes
place on the 1-d Hubbard string, the dynamics
of the string leads to spin charge separation 
in 2-d and restricted confinement as appears
in the form of stripes observed in experiment.
The ARPES data shows a significant Fermi
surface with hotspots, this cannot be easily
explained in terms of Fermi Liquid or arranged
non Fermi Liquid behaviour. The collective
dynamics of 1-d strings can provide a suitable
explanation. For example one can interpret that 
the holes Fermi surface are the regions where
the collective or luttinger liquid-like behaviour 
dominates. Let us try to see this point intuitively,
string theory has several conservation laws,
which follows due to various symmetries. Now
some of these symmetries are broken in the target
space, for example in context of gravity conformal 
invariance is broken in the target space by
gravity and preserved on the 2-d surface
generated by string and yet it is the condition
of conformal invariance on string generated surface
that gives Einstein's equation coupled to a
scalar field in the target space. Thus in an
analogous manner one may have obtain a situation
where Luttinger liquid symmetry is preserved
in some regions of the Fermi surface and badly
broken in others. As the Hubbard strings fluctuate
they generate a Fermi surface where Luttinger 
correlations are preserved in only some regions.
The area or volume where non-Fermi liquid
persists is expected to be proportional to
the number of strings, which is turn is
determined by physical parameters such as
doping etc. A challenging problem is to define
precisely a measure for the ``correlations'' in 
the system. 

In short we expect that by formulating
the underling theory in terms of collective
excitations such as strings [which
are indicative of Luttinger liquids] we can
represent a general theory for strongly
correlated systems in 2-d and 3-d.
Yet another support for our point of view comes
from data of Uemura, where he finds 2-d bose
like behaviour, and these fit reasonably with
Berezinskii-Kosterlitz-Thouless [BKT] type of phase 
transitions. Within our framework this is natural since 
conformal invariance is intimately tied with strings. 
It would be interesting to look into the details
of the relationship between Uemura's data and the 
BKT-type phase transitions. 

We now comment briefly on the choice of Hamiltoinan
and naively on one kind of duality in this context.  
	As already mentioned and as is well-known the 
t-J model is a special limit of Hubbard model. Keeping
this point in mind we want to seek a non-trivial
unification of Hubbard Hamiltonian and the t-J model
using `duality' of couplings in a new way. 
To achieve this unification we start with
a small step. Let us explain. To this end let us first 
recall the usual argument based on perturbation theory 
which allows us to see the t-J model as the special limit of
the Hubbard Hamiltonian [HH]. The HH reads
\begin{eqnarray}
H = -t \sum_{<ij>\sigma} c_{i\sigma}^{\dagger}c_{j\sigma}
+ U  \sum_{i} n_{i~\uparrow}  n_{i~\downarrow}, 
\label{d1}
\end{eqnarray} 
where $<ij>$ means sum over nearest neighbours, i.e.
hybridization between neigbouring atoms. $c_{i\sigma}^{\dagger}$
($c_{i\sigma}$) are the creation (annihilation) operators
of electrons at site $i$ of spin $\sigma$ ($\sigma \equiv \uparrow
, \downarrow$).
An important limit of HH is half-filled and strong coupling
, i.e. $U \gg t$. 

The intermediate state has one doubly occupied atom
and the effective interaction [second-order] is simply
\begin{eqnarray}
H = t \sum_{<ij>\sigma^{'}} c_{i\sigma^{'}}^{\dagger}c_{j\sigma^{'}}
\frac{-1}{U}t\sum_{\sigma} c_{i\sigma}^{\dagger}c_{j\sigma}, 
\label{d2}
\end{eqnarray}
which is the Hamiltonian of antiferromagnetically coupled
spin-$\frac{1}{2}$ Heisenberg model with coupling $J=2t^{2}/U$ 
per bond. Naively one can regard in some sense 
the Hamitonian in Eq.~\ref{d2} as a `dual' with respect
to the coupling $U$ of the Hamiltonian in Eq.~\ref{d2}
This argument is made in the context of degenerate
perturbation theory and the assumptions stated above. 
We now abandon any assumption of perturbation theory and 
simply assume to start with a Hamiltonian which is
the sum of the above two Hamiltonians, i.e.
\begin{eqnarray}
H &=& -t \sum_{<ij>\sigma} c_{i\sigma}^{\dagger}c_{j\sigma}
+ U  \sum_{i} n_{i~\uparrow}  n_{i~\downarrow}\nonumber\\
&&+ J \sum_{<ij>\sigma^{'}} c_{i\sigma^{'}}^{\dagger}c_{j\sigma^{'}}
\sum_{\sigma} c_{i\sigma}^{\dagger}c_{j\sigma}, \nonumber\\
J &=& - g^{2}/U,
\label{d3}
\end{eqnarray} 
where $g$ is some suitable coupling.

	In our scenario of HTSC theory we consider superconductivity 
arising from the dressing of 1-d stripe or string phase. 
The stripe arises due to a line of quantum critical points. This
is supported by Monte-Carlo calculations [among other
reasons] of 2D d-p Hubbard model where stripes and
superconductivity exists, see below. The correlation length in HTSC
material is short compared to their conventional cousins,
since here superconductivity arises due to 1-d stripes.
 
	In order to prove our conjecture we must consider
realistic Hamiltonians which exhibit stringy or stripe
solutions even in the superconducting state. One
such Hamiltonian is the 2d d-p model \cite{yan00},
\begin{eqnarray}
H &=& \varepsilon_{d}\sum_{i\sigma} d_{i\sigma}^{\dagger}d_{i\sigma}
+U_{d}\sum_{i} d_{i~\uparrow}^{\dagger} d_{i~\uparrow}  
d_{i~\downarrow}^{\dagger}d_{i~\downarrow}
\varepsilon_{p}\sum_{i\sigma} p_{i+\hat{x}/2,\sigma}^{\dagger}
p_{i+\hat{x}/2,\sigma} + p_{i+\hat{y}/2,\sigma}^{\dagger}
p_{i+\hat{y}/2,\sigma} \nonumber\\
&&+t_{dp}\sum_{i\sigma} \{ d_{i\sigma}^{\dagger}
(p_{i+\hat{x}/2,\sigma}+p_{i+\hat{y}/2,\sigma}
-p_{i-\hat{x}/2,\sigma}-p_{i+\hat{y}/2,\sigma})+h.c.\}\nonumber\\
&&+t_{pp}\sum_{i\sigma} 
\{
-p_{i+\hat{y}/2,\sigma}^{\dagger}p_{i+\hat{x}/2,\sigma}
+p_{i+\hat{y}/2,\sigma}^{\dagger}p_{i-\hat{x}/2,\sigma}
+p_{i-\hat{y}/2,\sigma}^{\dagger}p_{i+\hat{x}/2,\sigma}
-p_{i-\hat{y}/2,\sigma}^{\dagger}p_{i-\hat{x}/2,\sigma}
+h.c.\}
\label{d4}
\end{eqnarray}
For definitions of various terms see \cite{yan00}.
It is not easy to consider and clarify the ground state because
of the strong correlations between the d and p electrons.
The Variational Monte Carlo is used \cite{yan00} to examine
the overall structure of the phase diagram from weakly
to strongly correlated regions.
Some of the results of the MC calculations are summarized
in Fig.~\ref{fig1}-\ref{fig3}. Fig.~\ref{fig1} shows 
spin density and hole density. This is calculated the
for 16x16 d-p model.  The doping ratio is 1/8, U$_d$=8 and the 
level difference between d and p orbitals is 2.
In Fig.~\ref{fig2} the spin distribution in the ground state 
is given when the stripe is stable. The lengths of arrows
are proportional to the magnitudes of spins.
Finally in Fig.~\ref{fig3} the energy of stripe and commensurate 
SDW states is given.  The calculations are carried out on 
16x4 lattice at U$_d$=8 and the doping ratio=1/8. Squares denote 
the energy for commensurate SDW state, and circles and triangles indicate 
them for 4-lattice and 8-lattice stripes, respectively. 
Here for example, 4-lattice stripe means that there is
one stripe per 4 ladders. It is found in \cite{yan00}
that a picture of hole doping case which emerges from the 
MC evaluations is that the stripe state is stable at low doping and 
changes into the d-wave superconductivity. Thus we
can see from MC the underlying connection between
stripe [stringy] and superconductivity, which is a
numerical support of our conjecture, namely that 
superconductivity is related to stripes [i.e. 1-d structures] 
and/or can be understood in terms of such 1-d systems. It would
be interesting and useful to consider the electron-phonon
interaction in the 2D d-p model, and see if the
evaluations remain consistent with the current
results. 

	Yet another support for the validity of our
conjecture can be found in t'Hooft's idea of phase
transition in context of quark confinement\cite{hoo78}. 
It was shown by this author that in quantized gauge theories
one can introduce sets of operators that modify the
gauge-topological structure of the fields but whose
physical effect is in essence local. In 2+1 dimensions
it was shown for non-abelian gauge theories that these
operators form scalar fields, and when local gauge
symmetry is not broken spontaneously then these
topological fields develop a vacuum expectation
value and their mutual symmetry breaks spontaneously.
Given a gauge group SU(N) we have a non-trivial
center Z(N) of this group. Then one is lead to
the concept that topologically defined operators 
which create or destroy topological quantum
numbers can be thought of disorder parameters.
Here we again have a dual relation between
order and disorder. In most simple interpretation
a superinsulator is the dual of superconductor,
where the former is characterized by disorder.
	
	In conclusion we have indicated new ways of
looking at the physics of HTSC and related materials
and in general strongly correlated systems. Although
in its current form quantum groups are restricted to
1-d, we don't see this as a problem but a solution.
This 1-d restriction led us to think that we can
solve a 2-d or 3-d problem by using string theory 
as is done in case of gravity. Gravity is a 4-d 
phenomenon, but can be understood elegantly in 
terms of string theory [1-d objects]. This 
approach is fundamentally different from the one 
where one tries to go from 1-d to higher dimensions 
directly. Recent developments in string and topological
field theories can further help us to understand the 
physics of strongly correlated systems. In the context
of strongly correlated systems such as HTSC it
is the necessity to replace the Landau quasiparticle
by something else which can lead to a better
formulation and understanding of these sytems  
that has led us naturally to consider quantum
groups, strings and topology. We have also presented numerical 
support for our conjecture from variational MC evaluations.

\section*{Acknowledgments}
The Sher Alam's work is supported by the Japan Society for
for Technology [JST].


\begin{figure}
\caption{The spin density and hole density. This is calculated 
for 16x16 d-p model.  The doping ratio is 1/8, U$_d$=8 and the 
level difference between d and p orbitals is 2.}\label{fig1}
\end{figure}
\begin{figure}
\caption{The spin distribution in the ground state when 
the stripe is stable. The lengths of arrows
are proportional to the magnitudes of spins.}\label{fig2}
\end{figure}
\begin{figure}
\caption{The energy of stripe and commensurate SDW states.  
The calculations are carried out on 16x4 lattice at U$_d$=8 
and the doping ratio=1/8. Squares denote the energy for 
commensurate SDW state, and circles and triangles indicate 
them for 4-lattice and 8-lattice stripes, respectively. 
Here for example, 4-lattice stripe means that there is
one stripe per 4 ladders.}\label{fig3}
\end{figure}

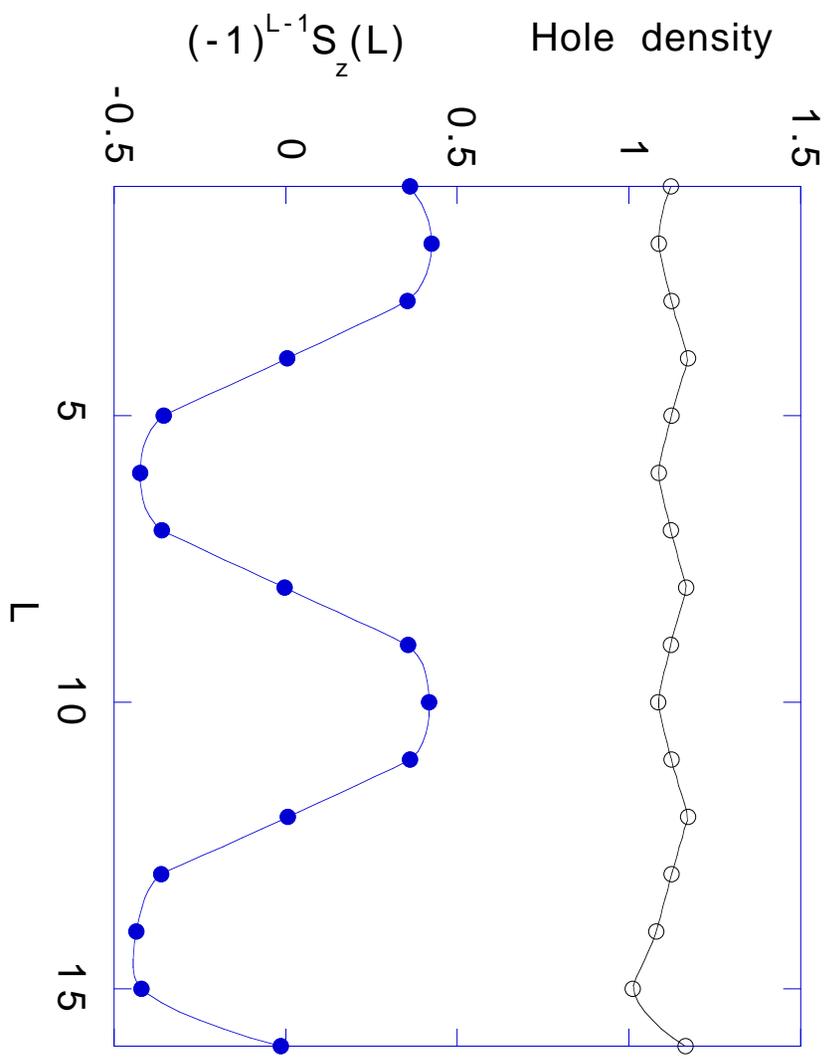
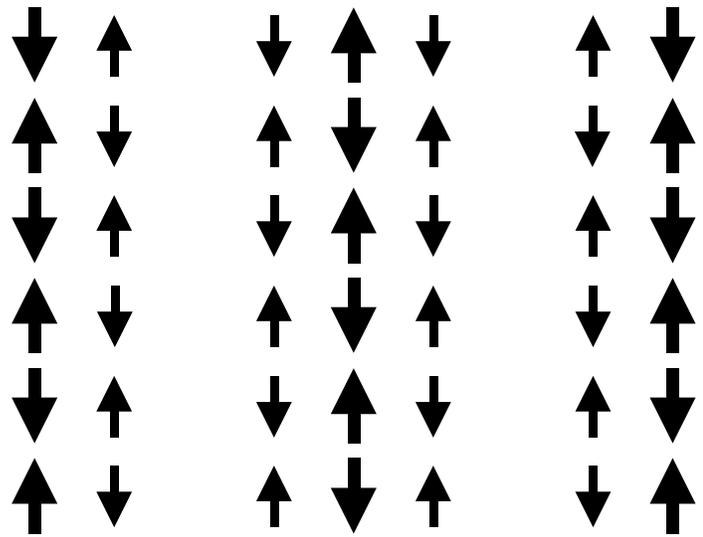
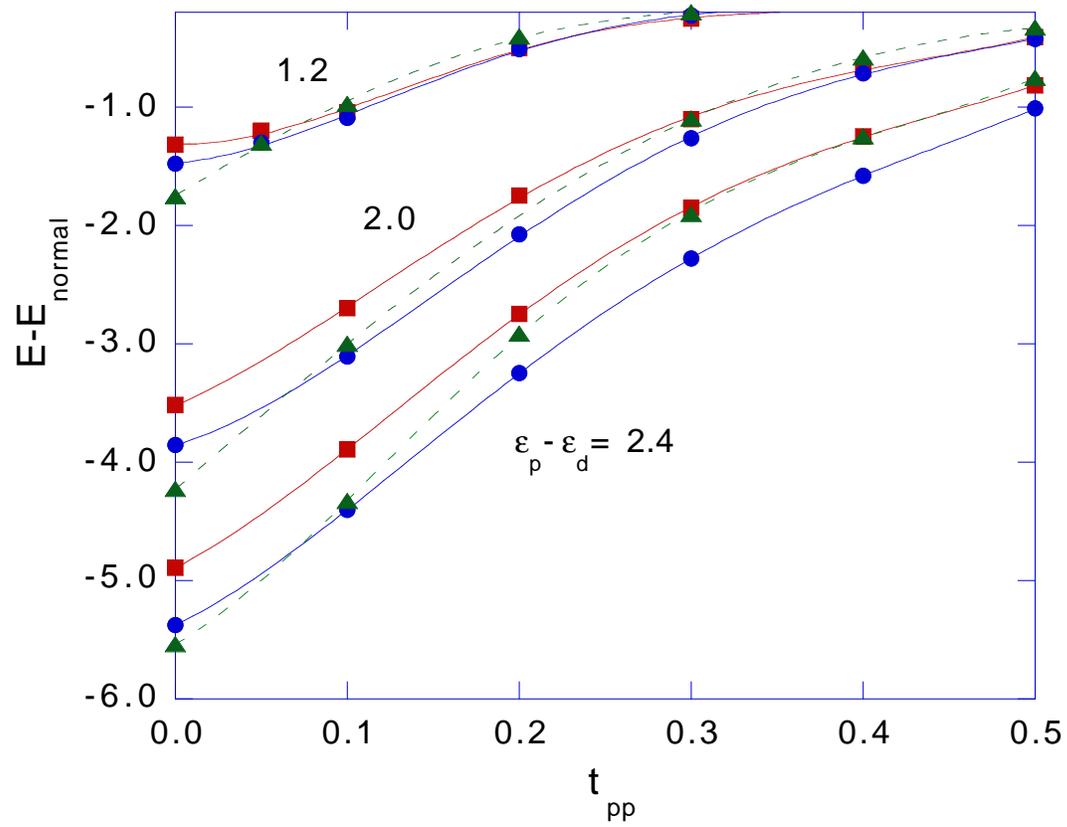
\end{document}